# Atomic-scale 3D imaging of individual dopant atoms in a complex oxide

K. A. Hunnestad[1], C. Hatzoglou[1], Z. M. Khalid[1], P. E. Vullum[2,3], Z. Yan[4,5], E. Bourret[5], A. T. J. van Helvoort[2], S. M. Selbach[1] and D. Meier[1,*]

[1]Department of Materials Science and Engineering, Norwegian University of Science and Technology (NTNU), 7491 Trondheim, Norway

[2]Department of Physics, Norwegian University of Science and Technology (NTNU), 7491 Trondheim, Norway

[3]SINTEF Industry, 7034 Trondheim, Norway

[4]Department of Physics, ETH Zurich, Zürich, Switzerland.

[5]Materials Sciences Division, Lawrence Berkeley National Laboratory, Berkeley, CA, USA

*dennis.meier@ntnu.no

**A small percentage of dopant atoms can completely change the physical properties of the host material. For example, chemical doping controls the electronic transport behavior of semiconductors and gives rise to a wide range of emergent electric and magnetic phenomena in oxides. Imaging of individual dopant atoms in lightly doped systems, however, remains a major challenge, hindering characterization of the site-specific effects and local dopant concentrations that determine the atomic-scale physics. Here, we apply atom-probe tomography (APT) to resolve individual Ti atoms in the narrow band gap semiconductor $ErMnO_3$ with a nominal proportion of 0.04 atomic percent. Our 3D imaging measures the Ti concentration at the unit cell level, providing quantitative information about the dopant distribution within the $ErMnO_3$ crystal lattice. High-resolution APT maps reveal the 3D lattice position of individual Ti atoms, showing that they are located within the Mn layers with no signs of clustering or other chemical inhomogeneities. The 3D atomic-scale visualization of individual dopant atoms provides new opportunities for the study of local structure-property relations in complex oxides, representing an important step toward controlling dopant-driven quantum phenomena in next-generation oxide electronics.**

The engineering of electronic responses with dopant atoms is crucial for modern technology. The functionality of diodes and transistors, for example, relies on semiconductors where dopant atoms generate the free holes (p-type) or electrons (n-type) that define the transport properties. Despite their substantial impact on the conductivity, the number of dopant atoms is usually small and even highly doped silicon contains just 1 dopant atom per ~$10^3$ Si atoms. In a more recent development, oxide-based semiconductors moved into focus as a particularly promising class of tunable systems for device applications. Analogous to conventional semiconductors[1], very low concentrations of dopant atoms can lead to pronounced changes in the electronic properties of oxide materials. The latter is reflected by doping-dependent studies on hexagonal manganites, where doping with aliovalent cations with a concentration of about 0.04 atomic percent (at. %) resulted in an order of magnitude lower electrical conductivity[2]. Furthermore, in complex oxides strong correlations between charge, spin and lattice degrees of freedom arise, promoting a wide variety of additional doping-induced effects, including insulator-metal transitions[3], interfacial magnetism[4] and superconductivity[5].

In contrast to more than 70 years of research on conventional semiconductors, however, the incorporation of dopant atoms in complex oxides is much less explored. Importantly, because of the symmetry reduction and strong electronic correlations, individual dopants can do much more than only control the type and concentration of mobile charge carriers. For example, dopants can induce local strain and strain gradients, electrostatic fields, orbital reconstruction, and novel magnetic phases[6]. Furthermore, dopants may occupy different regular lattice or interstitial sites with drastically different consequences for the physical properties of the host material[7,8]. In order to master this level of complexity and understand emergent composition-driven phenomena and opportunities in oxide materials, a careful characterization of the dopant atoms is essential. For this purpose, different experimental techniques, such as impedance spectroscopy[9], Hall measurements[10] and secondary ion mass spectrometry[11] have been applied, sensing average doping levels down to parts per billion. Despite their high sensitivity, these measurements cannot be applied to probe small volumes, let alone the lattice position of individual dopants and their interactions at the local scale, as they lack the necessary spatial resolution. To image single dopant atoms within the lattice and quantify their concentration, scanning transmission electron microscopy (STEM) was applied in combination with energy-dispersive X-ray spectroscopy (EDX)[12,13]. This correlated approach represented a breakthrough in the atomic-scale characterization of doped oxides, but it is limited to doping levels higher than a few at. %. Furthermore, the approach is inherently restricted to 2D projections along specific zone axes, prohibiting the full three-dimensional (3D) characterization of dopant atoms.

Here, we overcome this fundamental limitation by applying atom probe tomography (APT) to study the incorporation of dopant atoms in the narrow band gap semiconductor $Er(Mn,Ti)O_3$ ($E_g \approx 1.6$ eV[14]). By performing APT experiments on multiple needle-shaped specimens, we quantify the doping level and spatial dopant distribution, revealing substantial deviations of up to ≈ 50 % from the nominal dopant concentration as defined by the chemical stoichiometry during synthesis. Going beyond this quantitative analysis, we measure the 3D lattice position of individual dopant atoms, providing experimental proof that Ti is located within the Mn layers with no signs of chemical inhomogeneities.

APT is well-established in metallurgy, where it is widely used for element-specific 3D imaging, offering an unchallenged combination of chemical accuracy and sensitivity[15]. A key development that allowed for expanding APT studies towards a wider range of materials, including poorly conducting and even insulating systems, was the advent of laser-assisted field evaporation. Intriguing recent examples are APT experiments performed on frozen water[16], human enamel[17], multivariate metal-organic frameworks[18], and its application in geosciences[19]. Pioneering APT investigations on complex oxides were performed on $Pb(Zr,Ti)O_3$ ceramics[20] and $NiFe_2O_4$-$LaFeO_3$ nanocomposites[21], studying chemical composition and phase segregation, respectively. Atomically resolved APT data was achieved on $LiMn_2O_4$, showing the possibility to gain high-resolution 3D images of different lattice planes[22]. Furthermore, site-preferences of dopants were investigated in $AlGaO_3$ superlattices[23], but without locally resolving the dopants and lattice planes.

As a representative model material for the 3D imaging of individual dopant atoms in oxide-based semiconductors, we investigate $Er(Mn,Ti)O_3$ with a nominal Ti concentration of 0.04 at. %. The material belongs to the family of hexagonal manganites, $RMnO_3$ ($R$ = Sc, Y, In, or Dy to Lu), and exhibits a layered structure of Er atoms and corner shared $MnO_5$ bipyramids with $P6_3cm$ space group symmetry. The hexagonal manganites have been studied intensively with respect to their electric order[24], magnetism[25], and multiferroicity[26] and their basic physical properties are well understood. Enabled by the hexagonal crystal structure, the $RMnO_3$ family offers outstanding chemical flexibility and allows for doping to induce p- or n-type semiconducting properties at will. This flexibility was utilized in previous studies, controlling the density and type of the majority charge carriers via aliovalent cation substitution on both the A- and B-sites[2,27,28]. The applied characterization methods, however, did not allow for resolving the individual dopants. Instead, the lattice positions of dopant atoms were inferred from density functional theory (DFT) calculations. Due to inherent system size limitations, these calculations were conducted for substantially higher doping levels and without addressing effects that can arise during high-temperature crystal growth, such as dopant clustering, cation anti-sites, and non-stoichiometry[29].

To study the incorporation of Ti atoms into the $ErMnO_3$ crystal lattice, we begin with a basic atomic-scale characterization of $Er(Mn,Ti)O_3$. Figure 1(a) presents an illustration of the crystal structure that highlights the ratio of lattice atoms to dopants in our system, corresponding to one Ti atom in ≈ 2500 $ErMnO_3$ lattice atoms. One established method that can detect very low concentration levels of an atomic species is EDX.

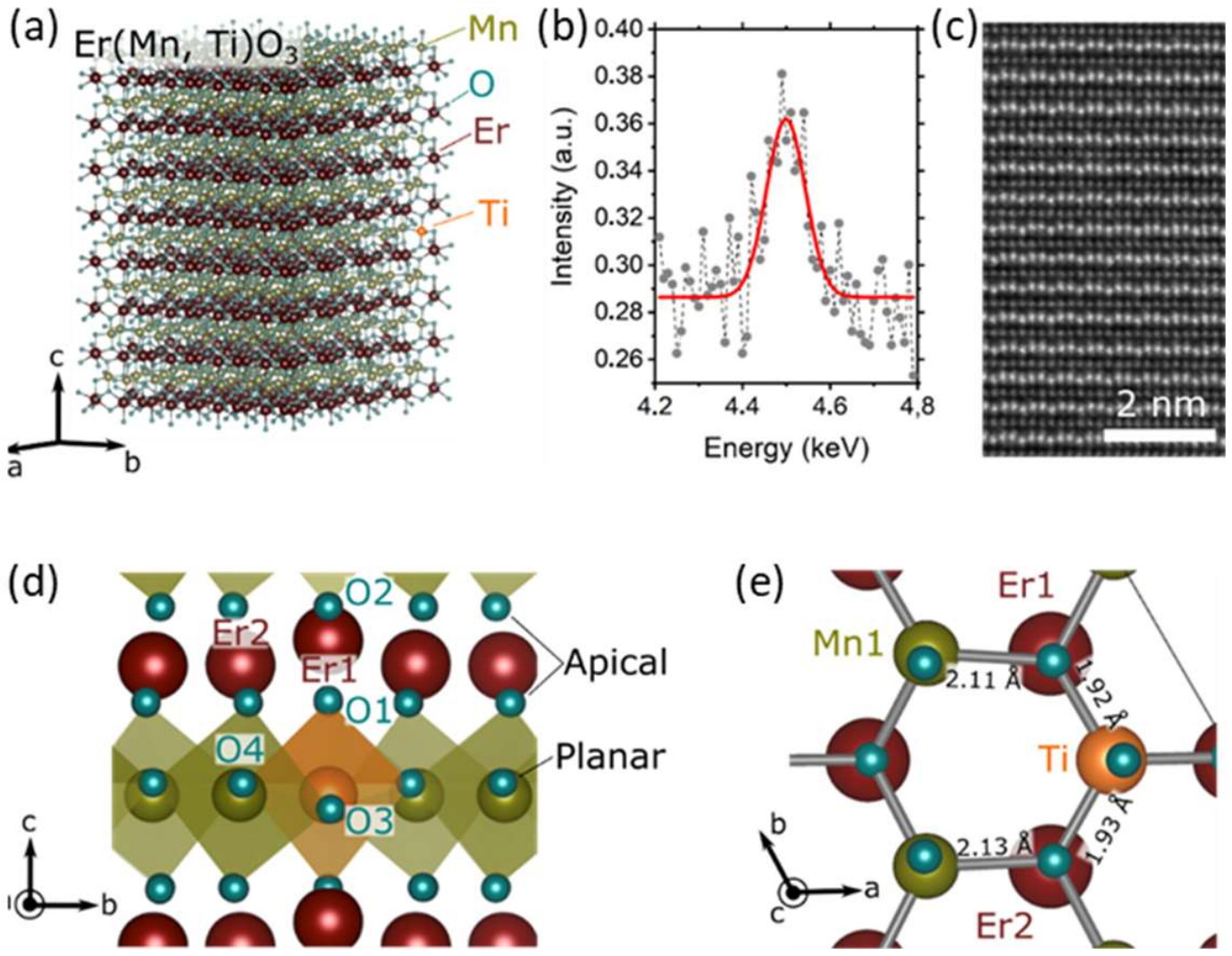

**Figure 1: Atomic-scale structure of Er(Mn,Ti)$O_3$.** (a) Illustration of the crystallographic structure of Er(Mn,Ti)$O_3$, showing the ratio between $ErMnO_3$ lattice atoms and Ti dopants. For a dopant concentration of about 0.04 at. %, there is approximately one Ti atoms per 64 unit cells of $ErMnO_3$. (b) Summed EDX spectrum from a 1.7µm line scan recorded on a Er(Mn,Ti)$O_3$ lamella. An X-ray peak is observed at 4.5keV, characteristic for the presence of the Ti K-edge. (c) HAADF-STEM image viewing along the *a*-axis gained from the same lamella as the EDX spectrum in (b). (d) and (e), Calculated local structure around the Ti dopant, situated on the B-site of the hexagonal $ErMnO_3$ system and viewed along the *a*- and *c*-axis, respectively.

A representative EDX spectrum of Er(Mn,Ti)$O_3$ obtained on a focused ion beam (FIB) cut lamella is shown in Figure 1(b). The spectrum represents the sum of about 900 spectra recorded in one line scan (≈ 1.7 µm) and reveals a peak at 4.5 keV, which is characteristic for Ti (the full spectrum is shown in Figure S1). The EDX data thus confirms the presence of Ti, but the measurement cannot provide the spatial resolution required to locate individual dopants. In order to resolve the structure of Er(Mn,Ti)$O_3$ at the atomic scale, we apply high-angle annular dark-field scanning transmission electron microscopy (HAADF-STEM) as shown in Figure 1(c) (see Methods for experimental details). The HAADF-STEM image in Figure 1(c) is recorded viewing along the *a*-axis of the same lamella as used for the EDX measurement in Figure 1(b), presenting the up-up-down arrangement of Er atoms typical for hexagonal manganites[30]. The layers with Er atoms are separated by Mn layers, which have a lower intensity in the HAADF-STEM image due to their lower atomic number Z. Although the crystal structure is readily resolved and atomic displacements can be probed with picometer precision, we observe no indication of the Ti doping in the HAADF-STEM data, consistent with the low concentration of dopant atoms and previous high-resolution measurements on Er(Mn,Ti)$O_3$[31,32]. To gain additional insight, we perform DFT calculations on a 540 atom 3x3x2 supercell of $ErMnO_3$ with 1 of 108 Mn atoms substituted with Ti (≈ 0.19 at. %), approaching a doping level comparable with the experimentally studied system. Substantial atomic displacements are only found for planar oxygen (O3 and O4 in Figure 1(d)) in the first coordination shell of the Ti dopant (Figure 1(e)). Compared to the undoped structure, the Ti-O bond length is shorter, whereas Mn-O bonds (Figure 1(e)) are subtly elongated, indicating partial reduction of adjacent Mn caused by $Ti^{4+}$ as a donor dopant in the $Mn^{3+}$ sublattice. Consistent with previous calculations for higher Ti doping levels[2], we observe that the strain field associated with the Ti atoms quickly decays (Figure S2), which makes them very difficult to observe in microscopy measurements as presented in Figure 1(c). In summary, the experimental data in Figure 1 shows that the applied Ti doping keeps the structural integrity at the nanoscale and the DFT calculations reveal the local structure in the ground state (T = 0 K) with $Ti^{4+}$ occupying the Mn sites. It is important to note, however, that Er(Mn,Ti)$O_3$ is synthesized at about 1450°C. At this temperature, both anions and cations are highly mobile and configurational entropy may favor, e.g., cation anti-sites and vacancies. This additional degree of complexity is not captured by the large-scale DFT model, reflecting the need for an experimental probe that can resolve the individual dopant atoms and, hence, clarify the atomic-scale structure.

In order to access the distribution of the dopant atoms in our Er(Mn,Ti)$O_3$ sample and gain spatially resolved data in 3D, we analyze the chemical atomic-scale structure using APT. For this purpose, needle-shaped specimens with a typical length of a few micrometers and a tip radius of less than 100 nm are extracted from a Er(Mn,Ti)$O_3$ single crystal using a FIB (see Methods for details of the extraction procedure). A representative SEM image of one of the FIB-cut needles used in our APT experiments is displayed in Figure 2(a). All extracted needles are oriented along the crystallographic *c*-axis as we confirm by electron diffraction (see Figure S10).

We begin the discussion of the APT results with the 3D distribution of the different atomic species in the analyzed volume. Figure 2(b) shows the APT reconstruction including all atoms, whereas only the Ti dopants are presented in Figure 2(c). The data suggests a homogeneous Ti distribution, which we confirm using radial distribution functions (RDF, Figure S3) and first nearest-neighbor analysis (1NN, Figure S4). The mass-to-charge ratio of the ions recorded during the APT field evaporation process is displayed in Figure 2(d) along with data recorded for undoped $ErMnO_3$. The mass spectra are centered around ≈ 79.9 Da, where a peak is observed only for Er(Mn,Ti)$O_3$. This peak is characteristic for the $TiO_2^+$ ionic species (see Figure S5 for the complete mass spectra and Figure S6 for the secondary Ti isotopes) and, combined with the $TiO^+$ ionic species, corresponds to a Ti concentration of 0.0086 at. %. In addition, Ti ions contribute to the $TiO^{2+}$ peak, which partly overlaps with the $O_2^+$ peak in the mass spectrum (Figure S7). In total, we thus find a Ti concentration of 0.0224 at. %, which translates into an average distance between Ti atoms of 2.08 nm for the investigated volume (Figure S4). To confirm that this concentration value is representative for our Er(Mn,Ti)$O_3$ crystal, we measure multiple APT needles from different locations as presented in Figure S8, revealing a Ti concentration of 0.0239 ± 0.0045 at.%, which is about 50 % lower than the nominal doping level as defined by the chemical stoichiometry during synthesis. Thus, the results in Figure 2(c) (and Figure S3 and S4) demonstrate a homogeneous Ti distribution, excluding clustering effects, chemical gradients, and other chemical inhomogeneities that may obscure the local electronic properties. Furthermore, we find that the Ti concentration is consistently lower than the nominal one, deviating by ≈ 50 % for the investigated region.

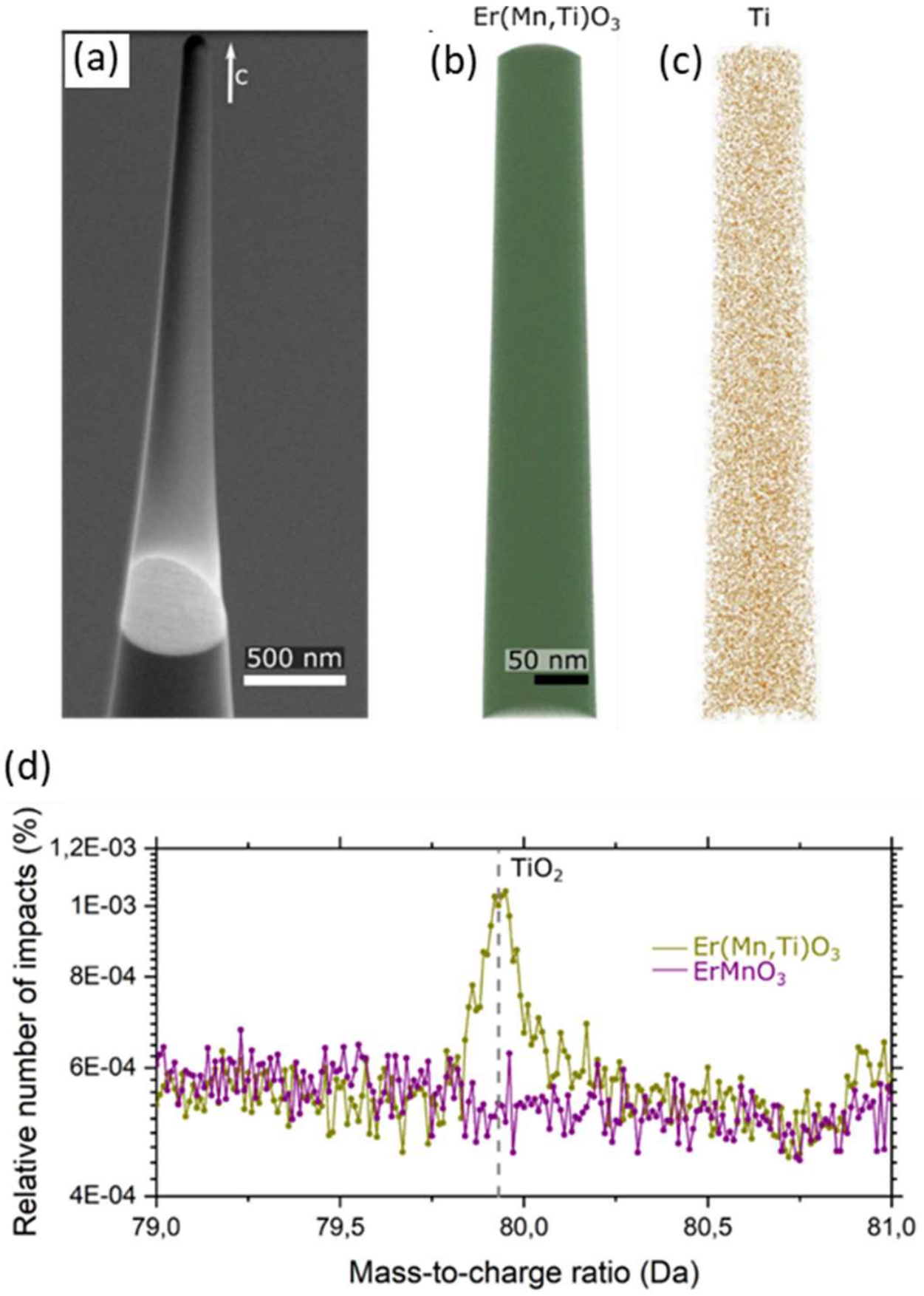

**Figure 2: SEM and APT images for a needle-shaped Er(Mn,Ti)$O_3$ sample.** (a) Scanning Electron Microscopy (SEM) image of a FIB-cut needle specimen with the crystallographic *c*-axis oriented along the needle axis. (b) APT dataset gained from the needle shown in (a), presenting all primary ionic elements (O – blue, Mn – yellow, Er – red, Ti – orange). (c) Subset of the APT dataset in (b), displaying only the dopants, i.e., $TiO_2^+$ ionic species. (d) Comparison of a selected section from the APT mass spectra recorded on Er(Mn,Ti)$O_3$ (yellow) and $ErMnO_3$ (purple). The Er(Mn,Ti)$O_3$ data shows a peak at ≈79.9 Da, which is characteristic for the $TiO_2^+$ species as indicated by the comparison with the spectrum of undoped sample.

Next, we consider the site-specific distribution of the different atomic species at the local scale, beginning with the $ErMnO_3$ lattice atoms (Figure 3). In the *c*-axis oriented needles, atomic planes of Er and Mn are readily resolved in the APT experiment as presented in Figure 3(a) and (b); Figure 3(c) displays the O atoms. To determine the atomic positions, we calculate spatial distribution maps (SDMs) in the evaporation direction *c* as shown Figure 3(d) (see Methods for details). The SDMs reveal the distance of the Er and O atomic planes relative to the Mn atoms measured along the crystallographic *c*-axis (Δz || *c*).

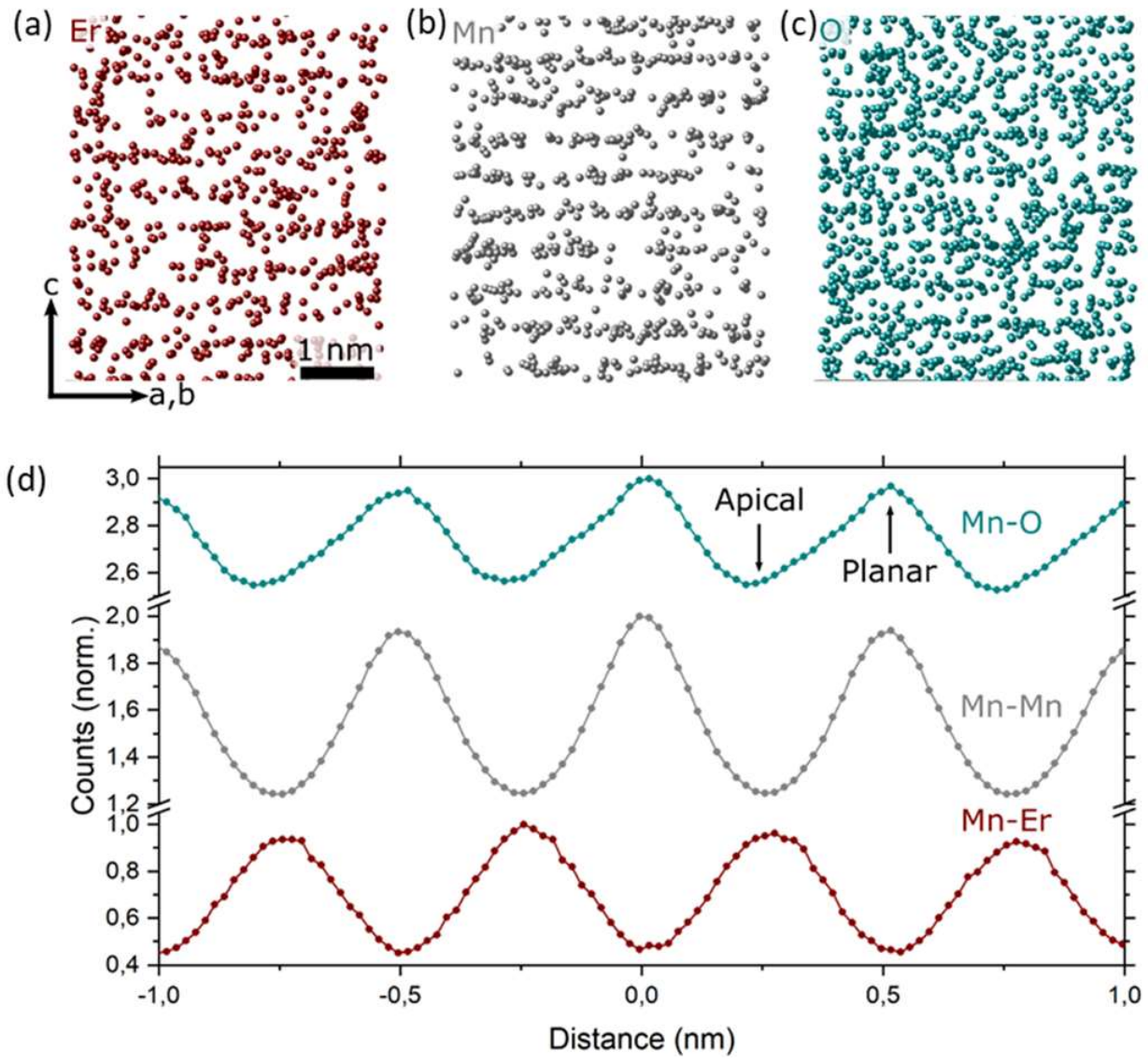

**Figure 3**: **3D imaging of the Er, Mn, and O lattice atoms in Er(Mn,Ti)$O_3$.** (a) APT reconstruction showing the Er lattice planes ($Er^{3+}$, $Er^{2+}$, $ErO^{2+}$ and $ErO^{+}$) from a region in the pole (see Figure S9). The volume is oriented so that the layered structure of Er atoms along the crystallographic *c*-axis is visible, consistent with the HAADF-STEM data in Figure 1(c). (b) and (c), Same as in (a) for Mn ($Mn^{2+}$ and $Mn^{+}$) and O ($O^{+}$) atoms, respectively. (d) SDMs of the lattice atoms, showing the distance between Mn and O (Mn-O) and Mn and Er (Mn-Er) with Mn serving as reference (Mn-Mn).

The Mn planes serve as reference system, so that Δz(Mn-Mn) = 0. We observe maxima in the counts for Δz(Er-Mn) at about ±0.25 nm with a periodicity of 0.51 nm, indicating that the Er planes are located between the Mn planes consistent with the HAADF-STEM image in Figure 1(c). Furthermore, the SDMs show that single O atoms (16 Da ionic species) are predominantly observed from the planar O positions, i.e., Δz(O-Mn) = 0, whereas apical O atoms are usually evaporated as $ErO_x$ species (Figure S5). In conclusion, Figure 3 shows that the 3D positions of the $ErMnO_3$ lattice atoms are readily resolved in the APT experiment, enabling a detailed analysis of the incorporation of the Ti dopants.

Following the same approach as for the $ErMnO_3$ lattice atoms, the position of individual Ti dopants is determined as presented in Figure 4, showing data from the same needle as evaluated in Figure 3. Figure 4(a) displays a volume from the [001]-pole, and Figure 4(b) and (c) show projections along and perpendicular to the crystallographic *c*-axis, respectively. Here, in addition to the Mn atoms (grey), the Ti atoms (or $TiO_2$ ionic species) are shown in orange. Importantly, the spatially resolved APT data indicates that the Ti dopants are located within the Mn lattice planes consistent with the calculations in Figure 1 and previous zero-Kelvin DFT results for lower doping levels[2]. To gain statistically significant information on the lattice position of the Ti atoms, we calculate SDMs over the entire needle length (> 300 nm) as shown Figure 4(d).

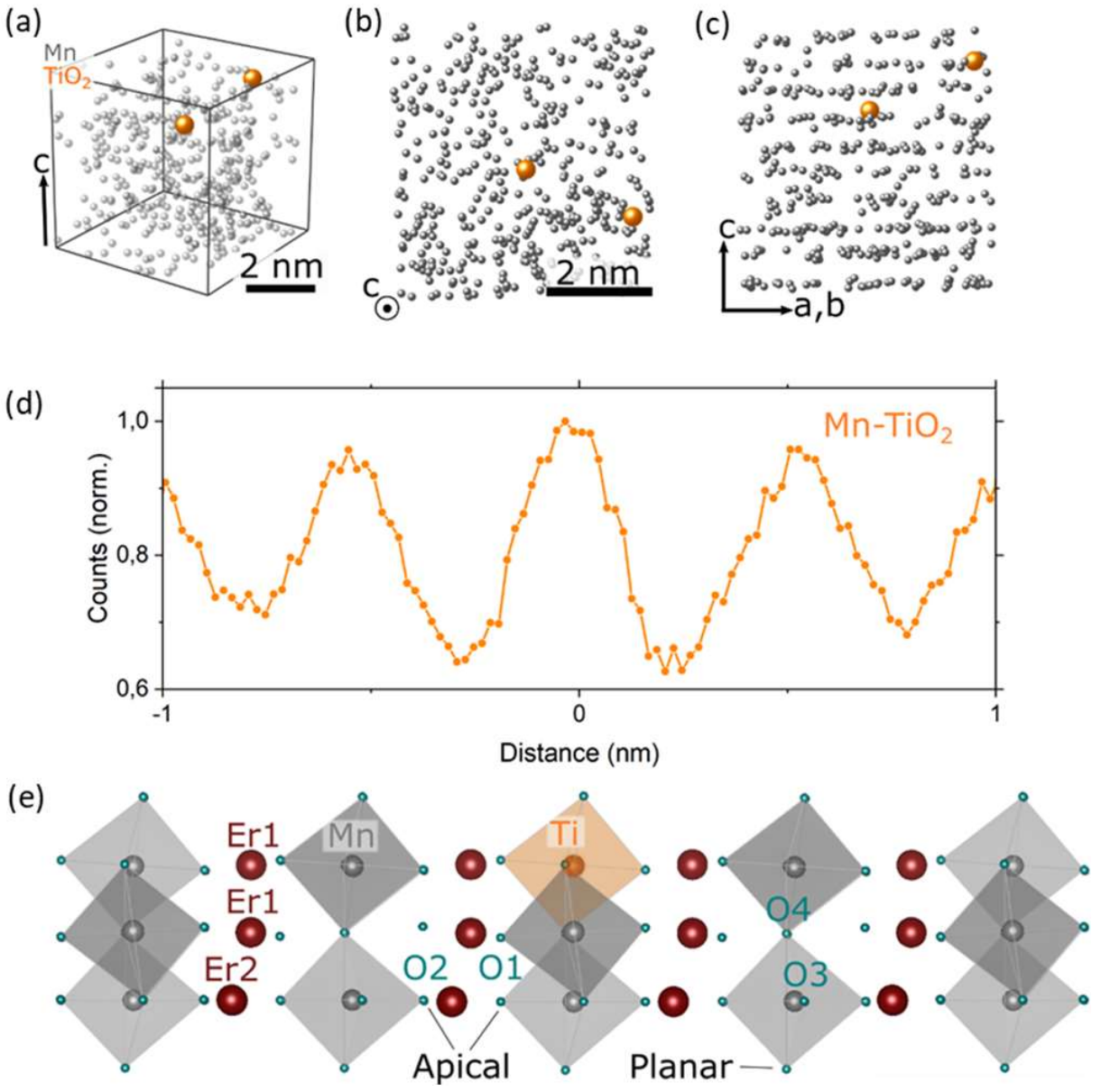

**Figure 4**: **3D Imaging of individual Ti dopant atoms in Er(Mn,Ti)$O_3$.** (a) APT reconstruction from a region in the pole presenting the Mn (semi-transparent grey) atoms and Ti (orange) atoms ($TiO_2^{+}$ species). (b) Same volume as shown in (a) viewed along the *c*-axis and (c), perpendicular to the *c*-axis so that the layered structure of the Mn atoms is visible. The data indicates that the Ti atoms are located within the Mn layers. (d) Spatial distribution maps (SDM) revealing the distance between of the Ti ions ($TiO_2^{+}$ species) and the Mn ions ($Mn^{+}$ and $Mn^{2+}$) in orange, with the latter serving as reference systems. (e) Schematic of the Er(Mn,Ti)$O_3$ crystal structure which summarizes the APT results, illustrating that the Ti dopant atoms are incorporated on B-sites of the hexagonal $ErMnO_3$ system.

Using the Mn layers as reference, we find a peak for Δz($TiO_2$-Mn) = 0, which corroborates that the Ti atoms are located in the Mn lattice planes. Figure 4 thus provides direct experimental evidence that the Ti dopants are replacing Mn atoms, occupying sixfold coordinated B-sites in the $ErMnO_3$ crystal structure as illustrated in Figure 4(e).

Our APT-based analysis can readily be expanded towards other complex oxides and atomic species, expanding investigations of dopant atoms into the realm of atomic-scale 3D imaging. The 3D imaging allows for quantifying otherwise elusive doping levels and determining site-preferences of dopant atoms in chemically complex host materials. The latter is particularly important for oxides as they are usually synthesized at elevated temperature, where configurational entropy promotes the formation of lattice defects. In addition, thermal gradients and cooling rates can play a crucial role for the concentration and distribution of dopant atoms, which makes it difficult to adequately model them, requiring atomically resolved measurements in combination with high chemical accuracy and sensitivity. Aside from the imaging of dopant atoms in semiconductors, the APT approach can be applied to topological insulators and superconductors to understand the impact of individual dopant atoms.[7,8,28] Furthermore, it may shed light on doping-related magnetic phenomena, such as the Kondo effect[33] and order-disorder transitions in multiferroics[34], adding an additional dimension to the atomic-scale investigation of emergent phenomena in complex oxides.

## Methods

**Sample preparation and characterization:** High-quality $ErMnO_3$ and $ErMn_{1-x}Ti_xO_3$ (x = 0.002) single crystals were grown by the pressurized floating-zone method[29], oriented by Laue diffraction and then cut to achieve oriented surfaces with the surface normal, *n*, parallel to the crystallographic *c*-axis. Samples were lapped and polished using silica slurry (particle size 20 nm) to obtain flat surfaces with sub-nanometer surface roughness. From the polished surfaces, APT needles with a tip radius $\lesssim$ 100 nm were prepared at 30 kV (and final polishing at 2 kV) using a Thermo Fisher Scientific G4 DualBeam UX Focused Ion Beam (FIB) analogous to the procedure described in Ref. [35]. Cross-section STEM samples were prepared using the same FIB. Carbon was used as protection layer on the regions of interest. The first part of the protection layer was deposited with electron beam assisted deposition to avoid $Ga^+$ implantation and beam damage into the top layer of the oxide. All coarse thinning was performed at 30 kV acceleration voltage for the ion beam. Final thinning was first done at 5 kV and finally at 2 kV on either side of the lamellae to minimize the surface damaged layer.

**Transmission electron microscopy and energy dispersive X-ray spectroscopy**: To analyze the sample quality and exclude, e.g., amorphization and structural damage from the FIB, selected specimens were inspected with STEM (Figure S10) using a JEOL 2100F Field Emission Gun (FEG) microscope, operating at 200kV. HAADF-STEM imaging of the lattice was performed on a double-Cs aberration corrected cold FEG JEOL ARM200FC at 200kV. Images were acquired with beam semi-convergence angles of 27.4 mrad and a beam current of 21 pA. The inner and outer semi-collection angles were 51 and 203 mrad, respectively. Energy dispersive X-ray spectroscopy (EDX) was performed with a 100 $mm^2$ Centurio detector, covering a solid angle of 0.98 sr. EDX line scans were performed with a 110 pA beam current and 1.0 s dwell time for each pixel. No visible beam damage could be observed after the EDX line scans. The EDX data were analyzed with DigitalMicrograph, version 2.32.

**APT data collection**: For the APT measurements a Cameca LEAP 5000XS was used, operating in laser mode. By applying laser pulses to temporarily heat up the specimen, the ions were controllably evaporated, and the time-of-flight was measured. Spatial positions were recorded with a 2D detector and determined by the electrostatic field lines of the specimen. The raw data consists of information of both time-of-flight, directly linked to the charge-to-mass ratio, and the detector impact position related to the original position on the tip prior to evaporation. The histogram (i.e., mass spectrum) and detector events for a representative APT measurement are shown in Figure S5 and Figure S9, respectively. Laser pulses with a frequency of 250kHz and energy between 2 and 30pJ were used. The detection rate was set between 0.5% and 2%, meaning that on average 5-20 atoms were detected every 1000 pulse. During the field evaporation, the specimen was cooled down to 25K.

**APT data reconstruction and analysis**: For the reconstruction of raw APT data into 3D datasets, the software Cameca IVAS 3.6.12 was used. Reconstruction was done in voltage mode with an image compression factor of ≈ 1.8, a field reduction factor of ≈ 2.8, and an evaporation field of Mn (30 V/nm). The parameters were fine-tuned using spatial distribution map (SDM) analysis so that accurate distances of atomic planes are measured in the reconstructed volume. For the peak at 16 Da which could correspond to either $O^+$ or $O_2^{2+}$, this is ranged as $O^+$ and not $O_2^{2+}$ following the discussion in Refs. [36,37].

**Spatial distribution maps (SDMs)**: SDMs are calculated by iterating through all the atoms, calculating the distance between atoms of one selected species and a reference species along a specific analysis direction. The distances are summed to generate a histogram as shown in Figure 3d[38]. To optimize the single-to-noise ratio, the analysis direction was chosen to be perpendicular to the atomic planes. The Mn-atoms (27.5 Da) are used as reference for all SDMs. For the matrix SDM, smaller volumes of about 10x10x10$nm^3$ were used, whereas for the dopants much larger volumes of 300x10x10$nm^3$ were required obtained from the [100]-pole region (Figure S9). Only the major single ionic species were used for the matrix, i.e., O from 16 Da ($1^+$ species), Mn from 27.5 Da ($2^+$ species), and Er from 55 Da ($3^+$ species). The Norwegian Atom Probe App (NAPA) software, developed in MATLAB ®, was used for all the SDM analysis.

**Density functional theory (DFT) calculations**: Our previous DFT calculations using 2×2×1 supercells showed that Ti in $ErMnO_3$ occupies the Mn site, acting as a donor and reducing the bulk p-type conductivity[2]. Here, we used 540 atom 3×3×2 supercells with a more realistic Ti dopant concentration of x=1/108 in Er(Mn,Ti)$O_3$. The calculations were done using VASP[39,40] with the projector augmented wave (PAW)[41] method and the PBEsol functional[42]. For Er, Mn, Ti and O, 9, 13, 12 and 6 electrons, respectively, were treated as valence electrons. Gamma point calculations were performed with a plane-wave cutoff energy of 550 eV and the geometry was relaxed until the forces on each ion were less than 0.03 eV/Å. A Hubbard $U$ of 5 eV was applied to Mn 3d states to reproduce the experimental bandgap and lattice parameters[2], and collinear frustrated antiferromagnetic order[43] was imposed on the Mn sublattice.

## Methods references

## Acknowledgements

The Research Council of Norway (RCN) is acknowledged for the support to the Norwegian Micro- and Nano-Fabrication Facility, NorFab, project number 295864, the Norwegian Laboratory for Mineral and Materials Characterization, MiMaC, project number 269842/F50, and the Norwegian Center for Transmission Electron Microscopy, NORTEM (197405/F50). K.A.H. and D.M. thank the Department of Materials Science and Engineering at NTNU for direct financial support. D.M. acknowledges funding from the European Research Council (ERC) under the European Union's Horizon 2020 research and innovation program (Grant Agreement No. 863691) and further thanks NTNU for support through the Onsager Fellowship Program and NTNU Stjerneprogrammet. Z.M.K. and S.M.S. thank the RCN for support through project 302506, and Uninett Sigma2 for providing computational resources through project NN9264K. Hanne-Sofie Søreide is thanked for her support to the APT lab facilities.

## Author contributions

K.A.H. prepared the APT samples using FIB, conducted the APT experiments and data analysis, supported by C.H. and under supervision from A.T.J.v.H. and D.M. Z.M.K. performed the DFT calculations supervised by S.M.S. P.E.V. performed the transmission electron microscopy measurements. Z.Y. and E.B. provided the materials. D.M. devised and coordinated the project and, together with K.A.H., wrote the manuscript. All the authors discussed the results and contributed to the final version of the manuscript.

# Supplementary Figures

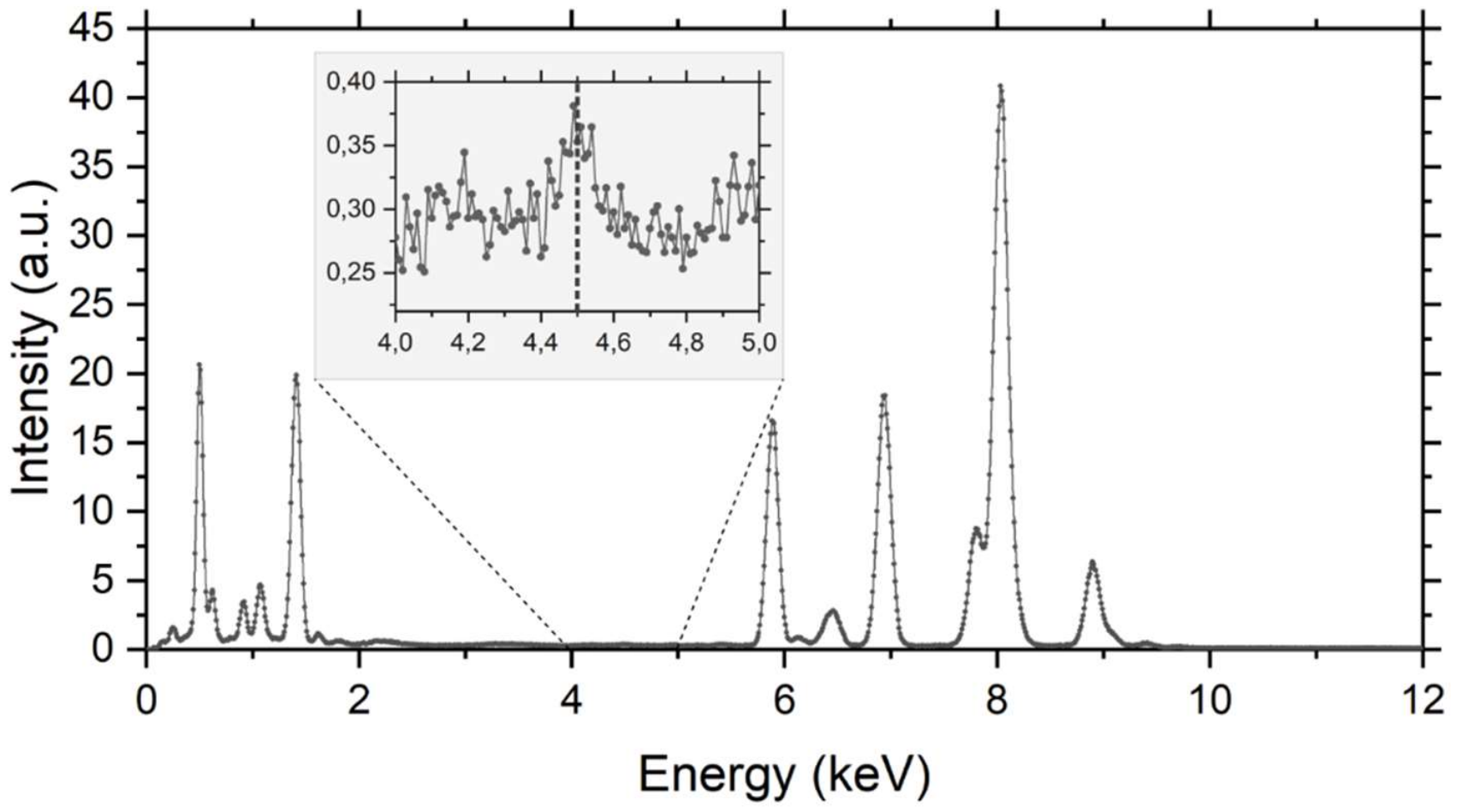

Figure S1: EDX spectrum of Er(Mn,Ti)$O_3$. The graph shows the complete EDX spectrum recorded on a Er(Mn,Ti)$O_3$ lamella (see main text and Methods for details). The inset shows the same part of the spectrum as Figure 1b in the main text with the Ti peak at 4.5 keV. The EDX spectrum represents the sum of $\approx$ 900 individual spectra recorded pixel-by-pixel along a single scan line with a length of 1.7 µm.

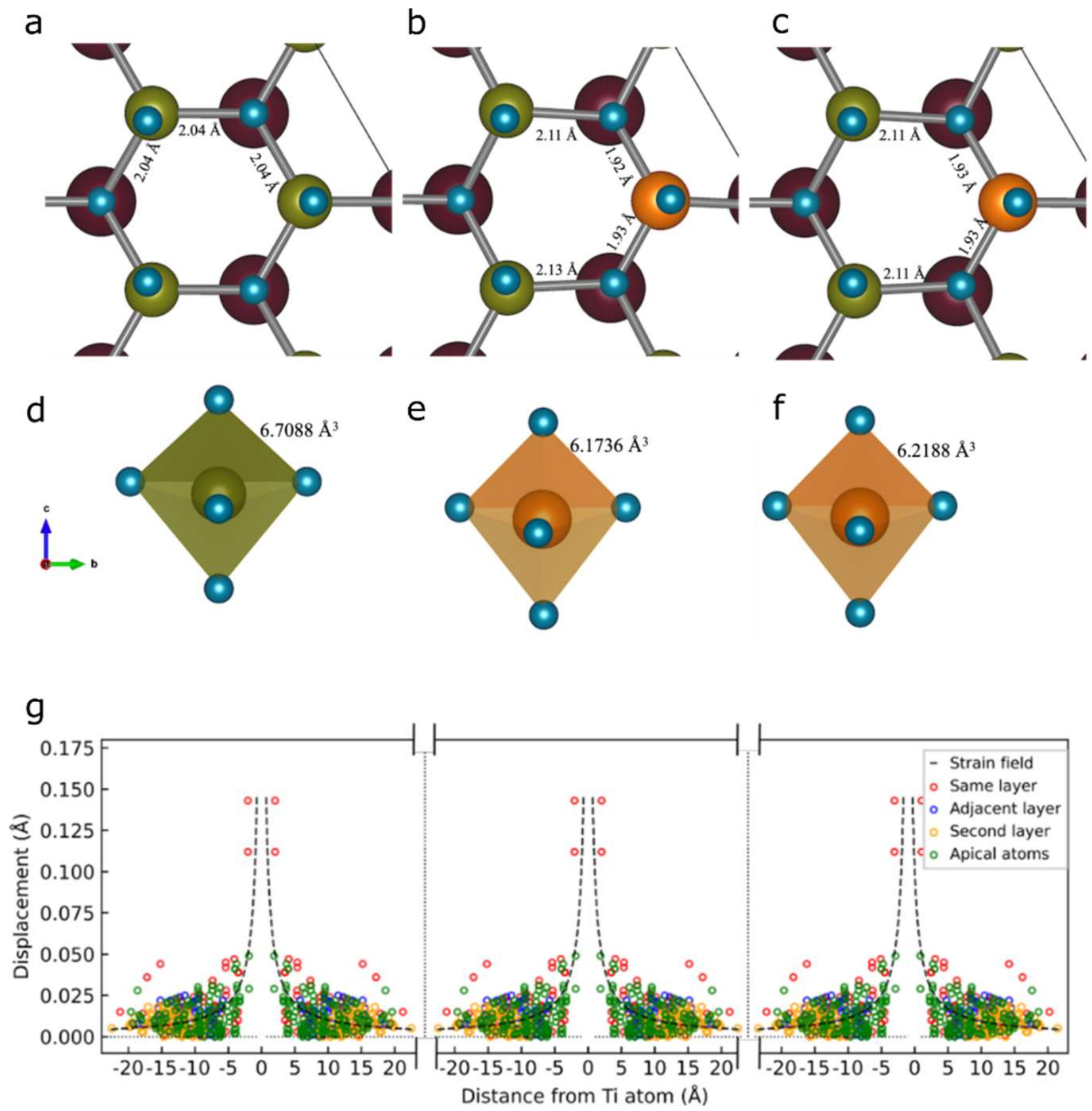

**Figure S2: Crystal perturbation upon Ti-doping.** The resulting supercell of $ErMnO_3$ for (a)**,** stoichiometric undoped structure (b), neutral supercell (c), positively charged (+1) supercell and (d)-(f) shows the corresponding expansion and contraction of the (Mn,Ti)$O_3$ polyhedral at the doping site. (g) Local strain field around Ti dopants visualized from DFT calculated ion displacements relative to the structure of undoped $ErMnO_3$ as a function of distance from a Ti dopant. Vertical dotted lines indicate boundaries between periodic supercells.

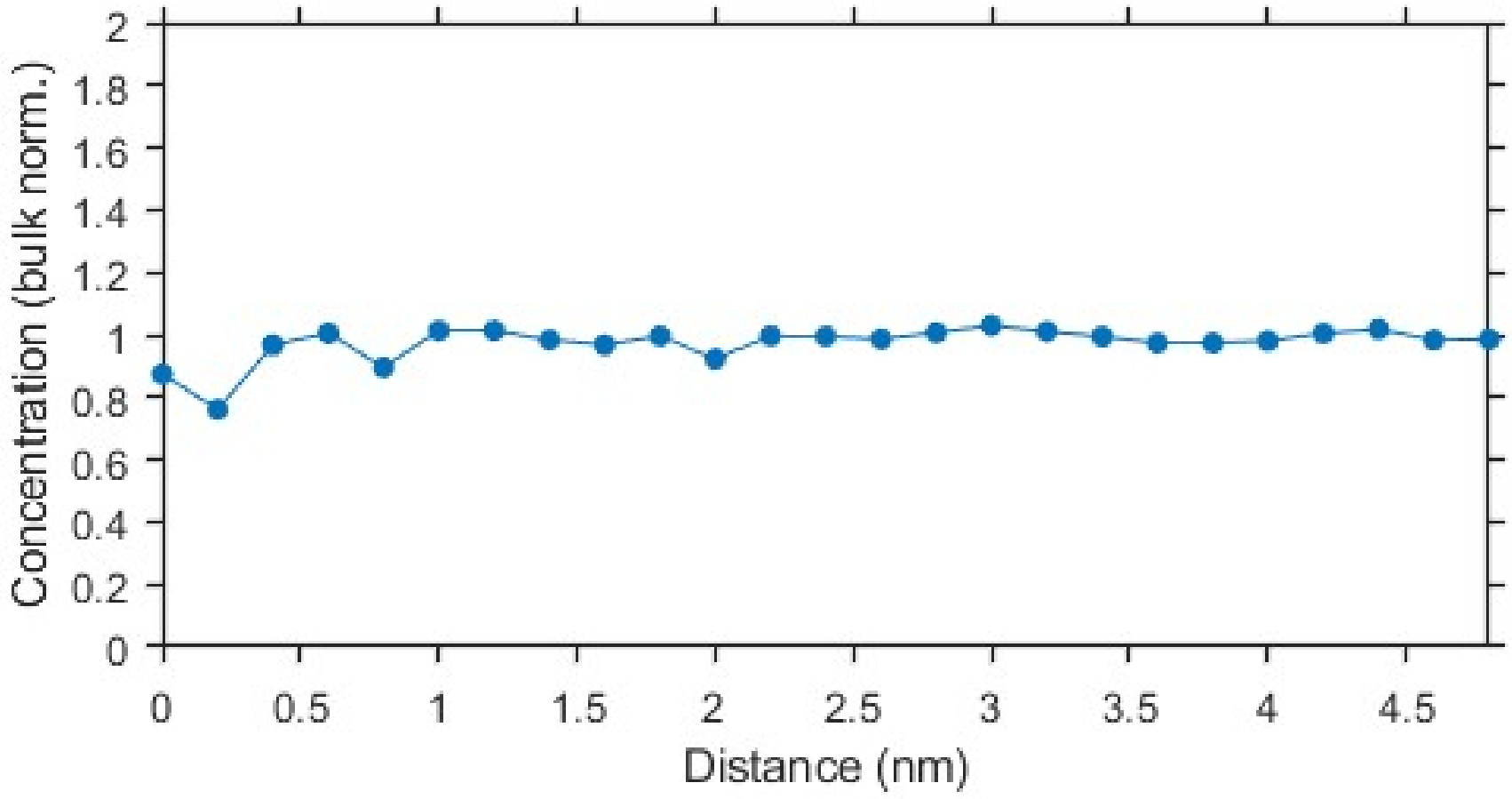

**Figure S3**: **Radial distribution function (RDF) analysis of Ti dopants.** The graph displays the RDF analysis for Ti dopants for the sample presented in Figure 2 in the main text. The calculation shows the chemical composition of Ti inside shells of increasing radii centered around the $TiO_2$ ions, normalized to the overall bulk composition, and displayed as a histogram. We note that only the $TiO_2^+$ ionic species was used as this was the most pronounced Ti species with the least amount of noise and background. The RDF analysis shows no substantial deviation from 1 at small distances, indicating a homogeneous distribution of Ti atoms.

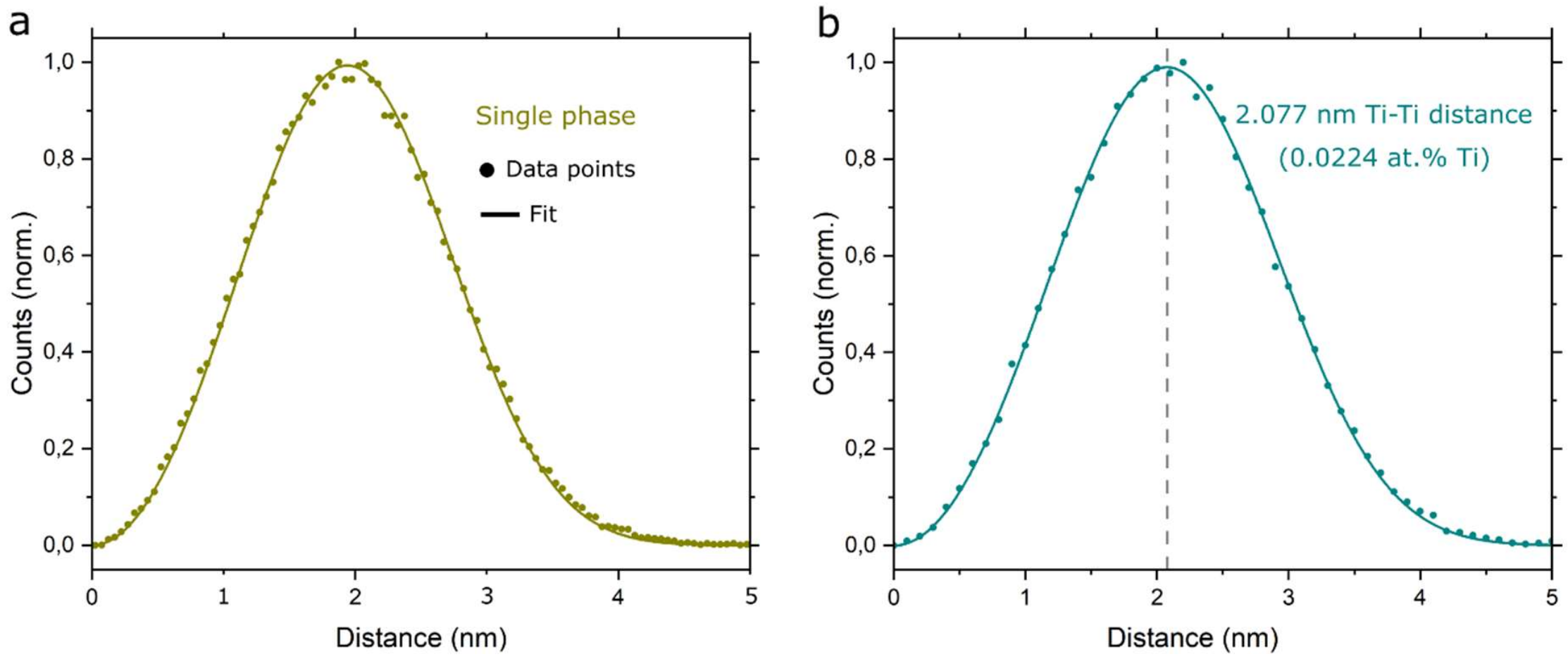

**Figure S4: First nearest neighbor (1NN) calculations for Ti dopants.** (a) 1NN calculations for the TiO and $TiO_2$ species based on the experimental data for the specimen shown in Figure 2 of the main text. The solid line is a fit to the APT data, based on the model for a single-phase and homogenous distribution in Ref. [1], which was used to study secondary phases or clustering. The x-axis gives the separation distance between neighboring ions, whereas the y-axis shows the number of ions with this separation distance normalized to the maximum value. It is important to note that the experimental curve cannot be used to extract quantitative information as detection efficiency and background signals are not taken into account. The excellent agreement between model and experiment, however, allows to exclude clustering of the Ti atoms and Ti-related secondary phases. The latter implies that the 1NN distance of the Ti dopants can be calculated from the measured concentration, i.e., 0.0224 at. % for the sample in Figure 2 of the main text as shown in (b)**,** corresponding to an average 1NN distance of 2.077 nm (solid line: fit to the data, same model as above). The deviation in 1NN distance between the experimental (a) and simulated (b) data is due to a combination of a relatively high background level and a limited detection efficiency (Figure S11), in addition to a different selection of ionic species ((a) uses only the $TiO_2^+$ species, while (b) is based on all species).

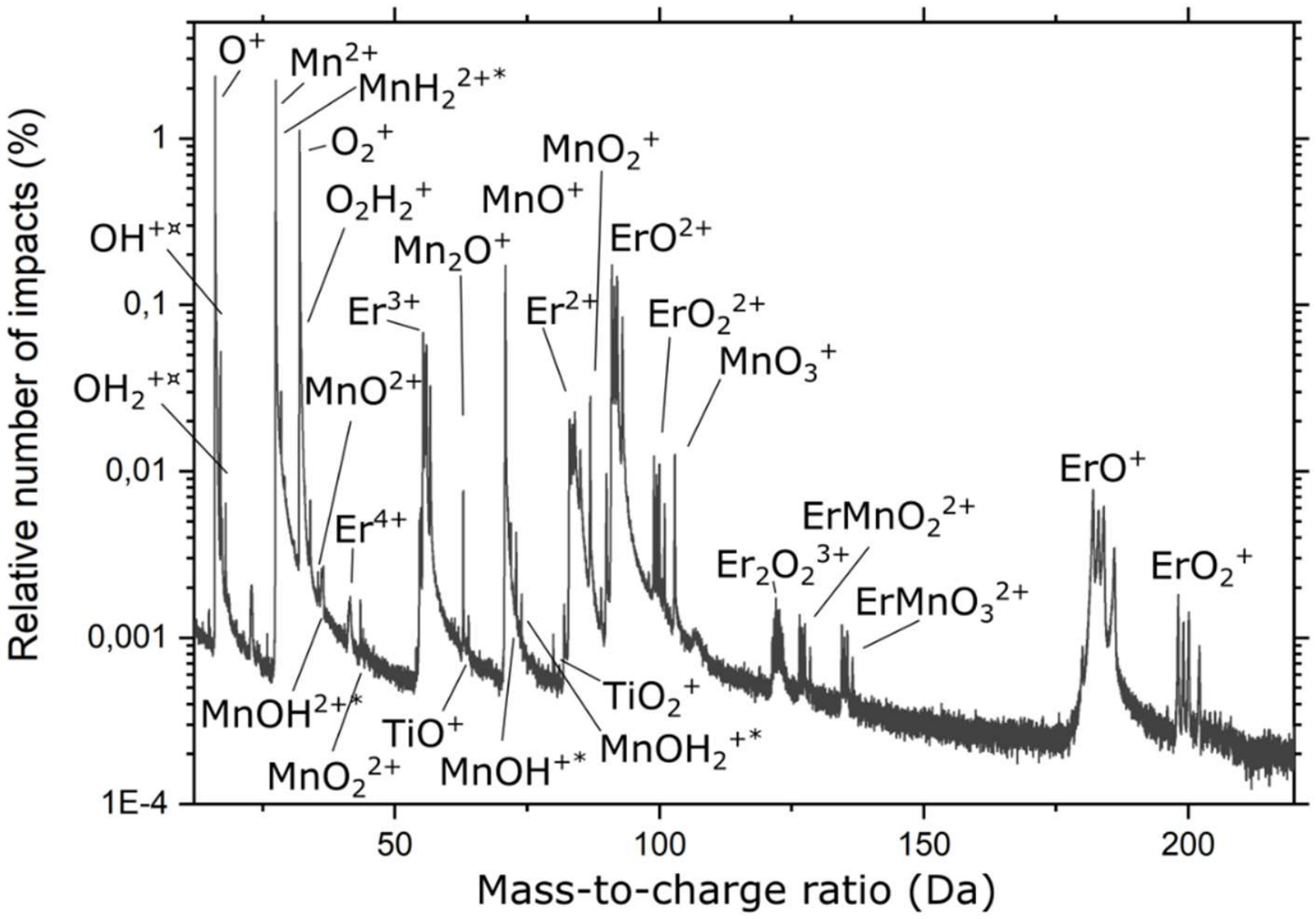

**Figure S5: Mass spectrum of $Er(Mn,Ti)O_3$ with labels for the ionic species.** The horizontal axis shows the mass-to-charge ratio obtained from the APT time-of-flight measurements. The y-axis displays the relative number of ion impacts on the detectors in logarithmic scale. *Denotes H-species where the H-atom is not included in the reconstruction, i.e., MnOH is counted as MnO. ¤Denotes species that do not originate from the sample, but from the APT chamber. Species below 14 Da most likely originate from the APT chamber and are therefore not shown.

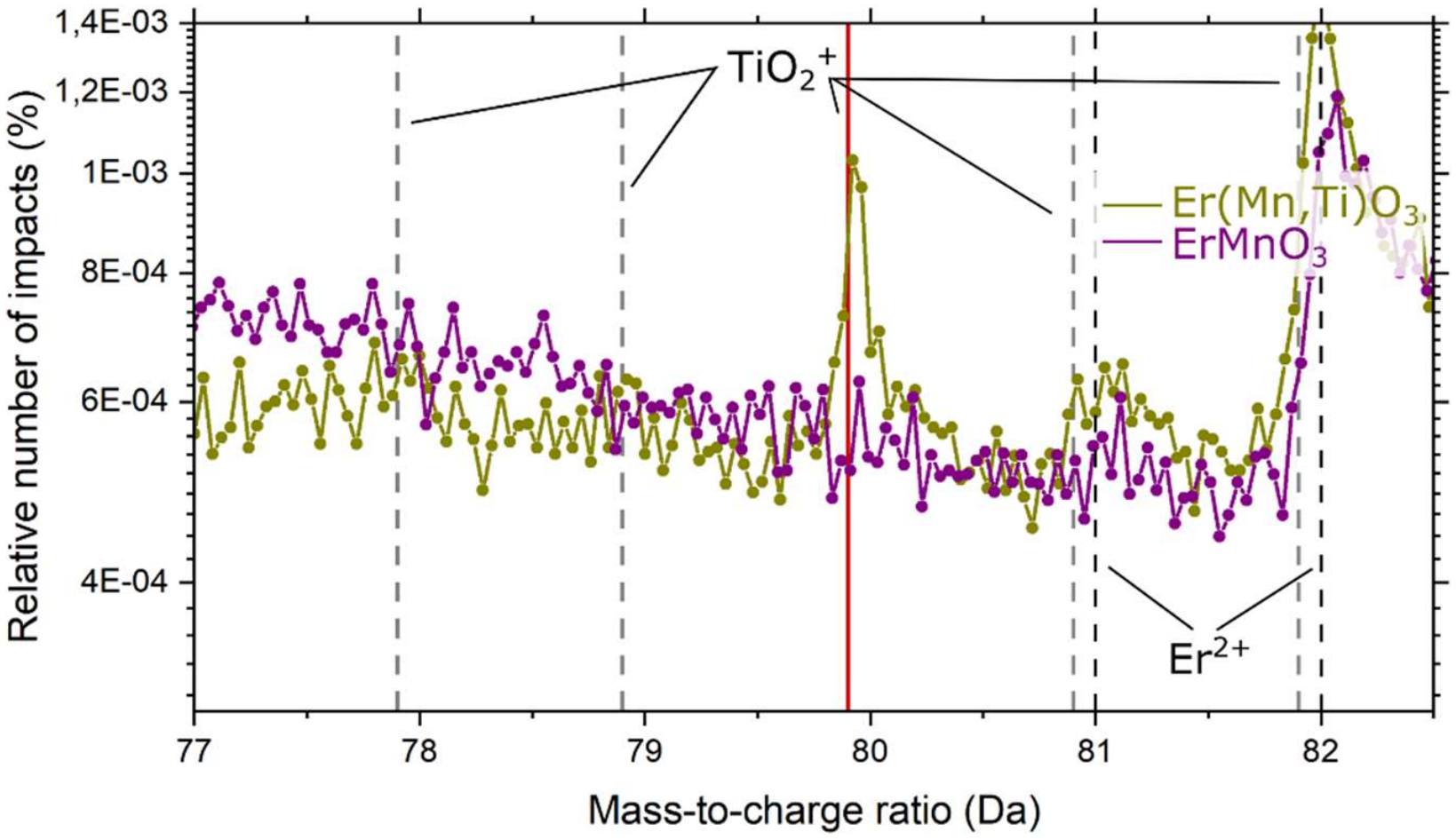

**Figure S6: Extended mass spectrum of $TiO_2^+$ ionic species**. The graph represents an extension of Figure 2d in the main text, where the visible peak of the ionic species of $TiO_2^+$ is shown, as well as the minor isotopes of $TiO_2^+$ (dashed gray lines). The latter are much less pronounced and partly overlapping with the $Er^{2+}$ ionic species.

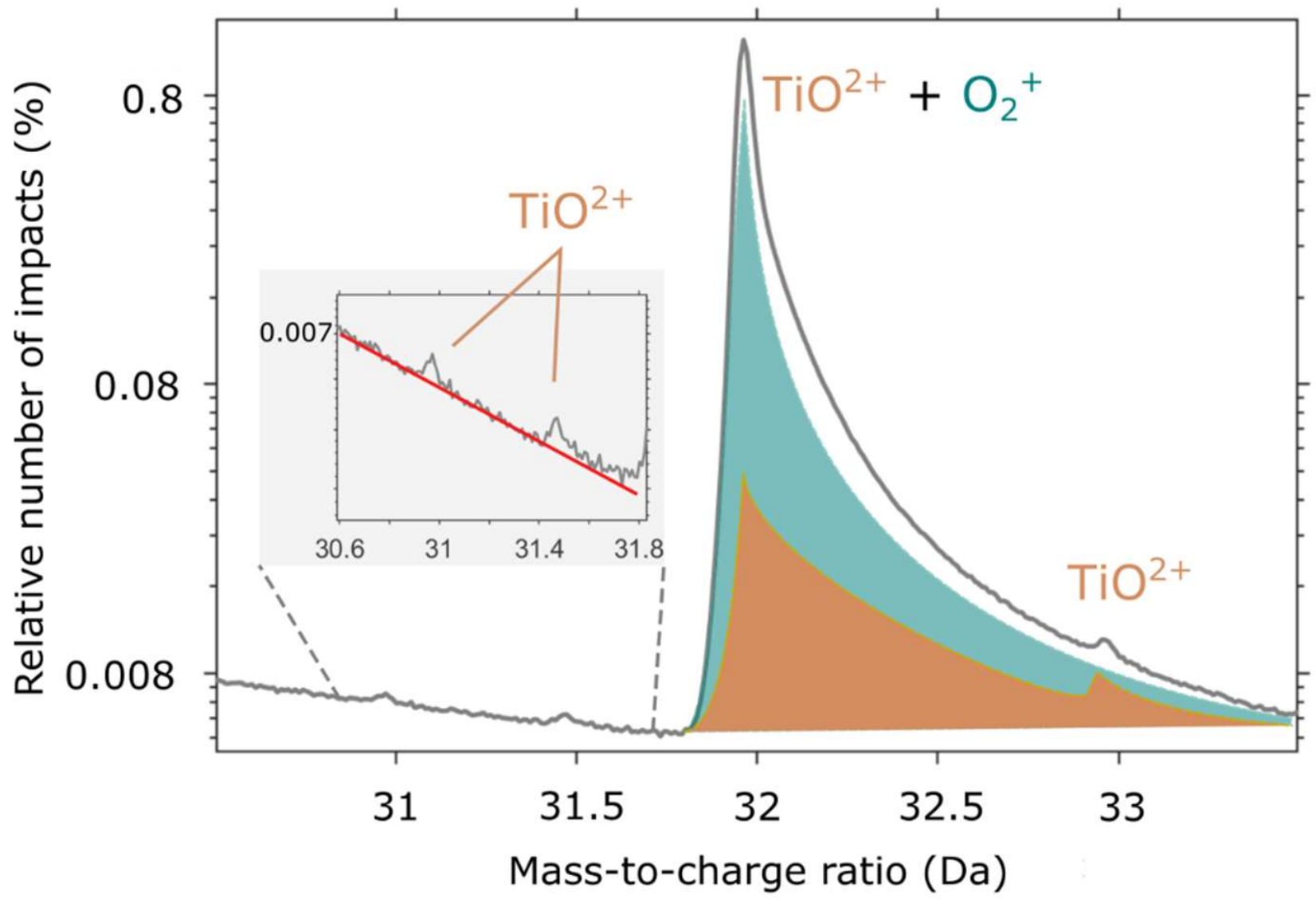

**Figure S7: Mass spectrum overlap of $O_2^+$ and $TiO^{2+}$.** Mass spectrum centered around the 32 Da peak which contains both the $O_2^+$ and $TiO^{2+}$ ionic species, as illustrated by the colored regions in the figure (orange is $TiO^{2+}$ and blue is $O_2^+$). The $O_2^+$ is by far most dominant and hides the main $TiO^{2+}$ peak, but the minor isotopes are visible. By using an exponential fit for the background (see inset, red line is illustrative), the total ion counts in the minor isotopes on the left can be used to deconvolute the main peak at 32 Da and extract an estimate for the $TiO^{2+}$ concentration which is found to be 0.0138 at.%. Adding this to the previously measured Ti concentration of 0.0086 at.% the total concentration becomes 0.0224 at.%.

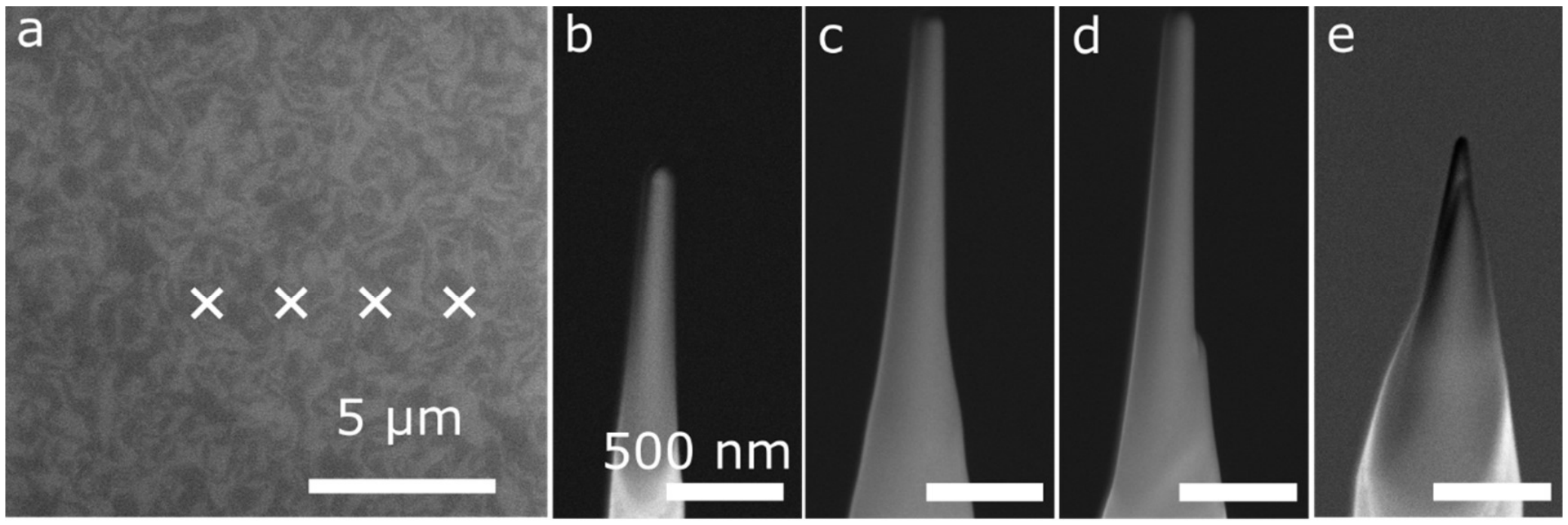

**Ti concentration (at.%)** 0.0224 ± 0.0001 0.0226 ± 0.0003 0.0201 ± 0.0003 0.0303 ± 0.0004

**Figure S8: Sample geometry and Ti concentration for different $Er(Mn,Ti)O_3$ needles.** To evaluate spatial variations in Ti concentration, specimens were extracted from different locations within a region of about 20µm. (a) SEM image of the bulk sample showing contrast from the ferroelectric domains (see Ref. [2]) and markers indicating typical distance between each of the needles in (b). Four specimens (b)-(e) were analyzed with identical APT parameters of 5 pJ laser pulse energy, 250 kHz laser pulse frequency, 50 K base temperature and 2% detection rate. Below the needles is the Ti concentrations and error estimates for each needle (based on counting statistics) resulting in an average concentration of 0.0239 ± 0.0045 at.%. The value reflects only smaller variations in Ti concentration over micrometer distances, with an error bar that is the standard deviation based on the four values given and is likely to be dominated by the deviations in the shape of the needles, which determines the electric field strength at the apex.

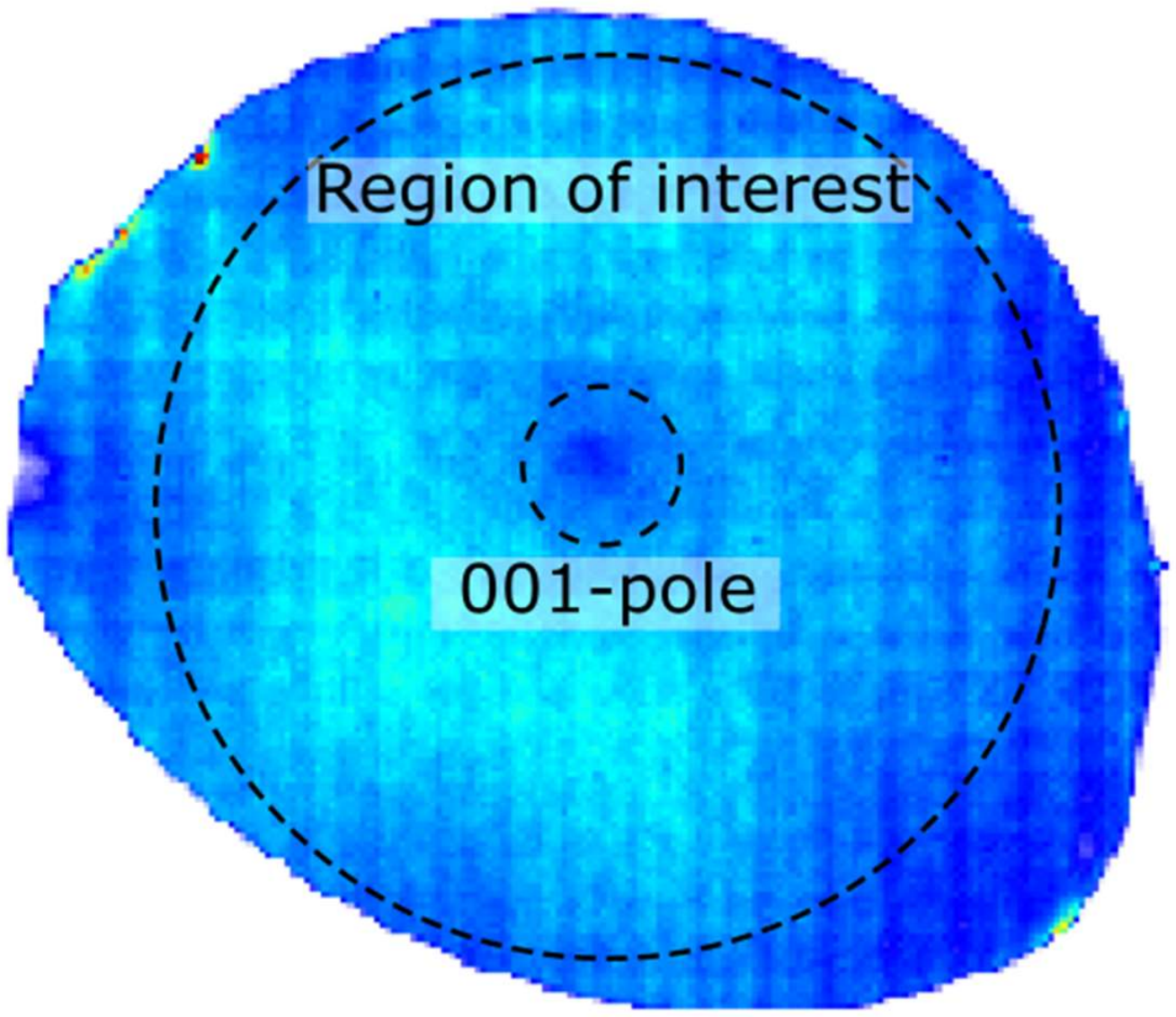

**Figure S9: Detector map.** The image shows the detector view after completing the evaporation process. The map contains all the detected hits and is displayed as an image where bright (red) indicate more hits than dark (blue). The full region of interest is used for the reconstruction and for analyzing the full volume of the APT needle discussed in the main text, while the 001- pole is used for obtaining atomic resolution (roughly corresponding to the dotted lines). This pole corresponds to a facet on the surface of the needle formed during the field evaporation.

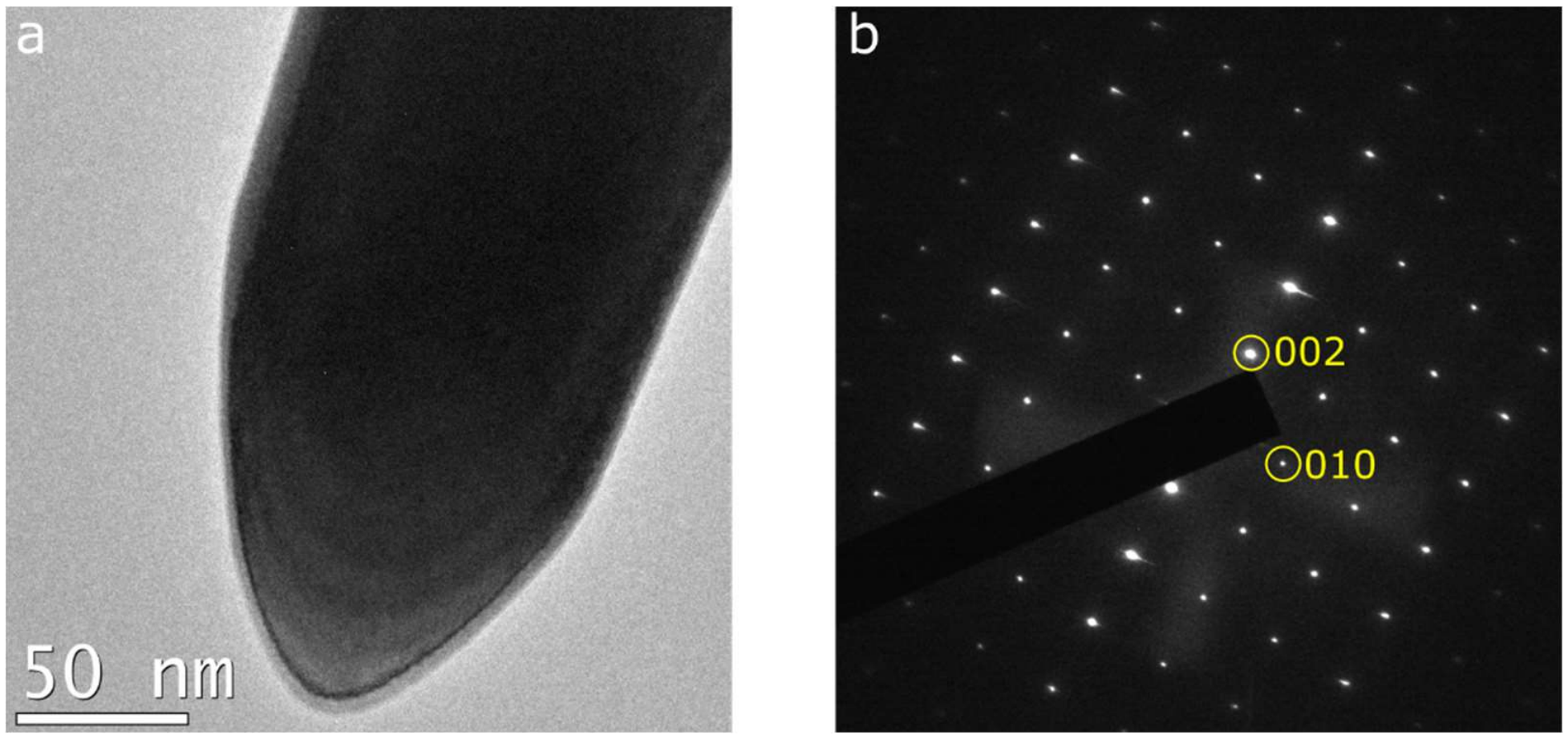

**Figure S10: Sample inspection with TEM.** (a) TEM image of the tip of a representative finished needle of $ErMnO_3$. The needle is slightly asymmetric, but no structural defects are observed and the bulk of the needle is crystalline. (b) Selected-area electron diffraction (SAED) pattern of the needle in (a). Only peaks associated to the hexagonal phase are seen, with no indication on secondary phases, reflecting the high-quality of the APT needle.

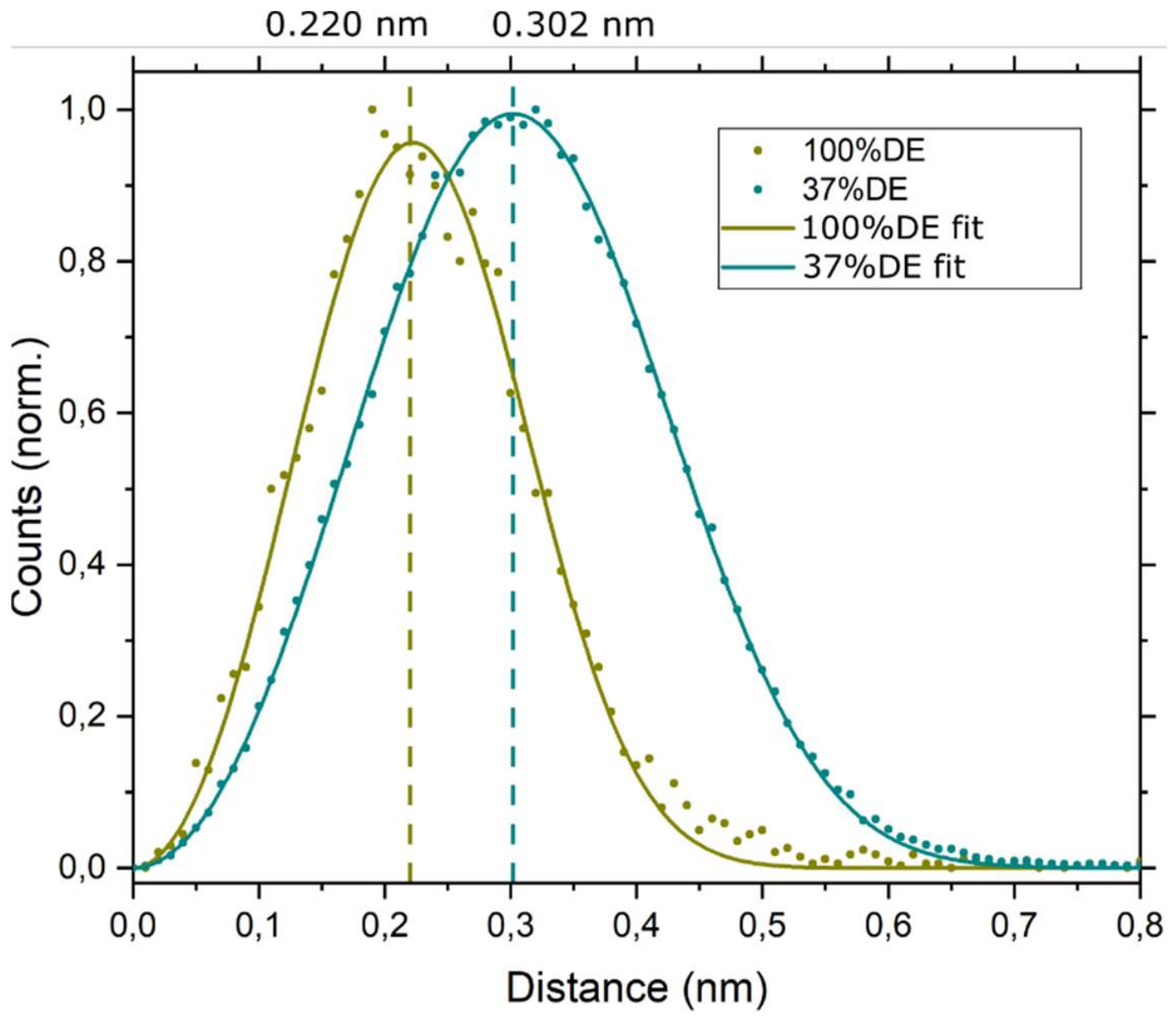

**Figure S11: Effect of detection efficiency on 1NN calculations.** Calculated 1NN of Er from simulated datasets of undoped $ErMnO_3$ at two detection efficiencies (DE); 100% (blue) and 37% (yellow). Both datasets are fitted (solid line) and the simulation with low detection efficiency is closely matching the experimental data, meaning 37% DE is close to the real experimental conditions. The center of the fit can be seen to be shifted from 0.220nm with 100% DE to 0.302nm with 37% DE.